\documentclass[sigconf, screen, authorversion]{acmart}

\usepackage{color}
\usepackage{enumitem}
\usepackage{threeparttable}
\newcommand{\e}[1]{\textcolor{blue}{#1}}

\begin{document}

\title{The Popularity Hypothesis in Software Security}
\subtitle{A Large-Scale Replication with PHP Packages}

\author{Jukka Ruohonen}
\email{juk@mmmi.sdu.dk}
\orcid{0000-0001-5147-3084}
\affiliation{
  \institution{University of Southern Denmark}
  \city{S\o{}nderborg}
  \country{Denmark}}

\author{Qusai Ramadan}
\email{qura@mmmi.sdu.dk}
\orcid{0000-0001-8159-918X}
\affiliation{
  \institution{University of Southern Denmark}
  \city{S\o{}nderborg}
  \country{Denmark}}

\begin{abstract}
There has been a long-standing hypothesis that a software's popularity is
related to its security or insecurity in both research and popular
discourse. There are also a few empirical studies that have examined the
hypothesis, either explicitly or implicitly. The present work continues with and
contributes to this research with a replication-motivated large-scale analysis
of software written in the PHP programming language. Two datasets are used: the
first contains nearly four hundred thousand open source software packages
written in PHP and the second addresses over six thousand WordPress
components. According to the results based on vulnerabilities reported, the
hypothesis holds: packages having seen reported vulnerabilities over their
release histories are generally more popular than packages for which fewer or no
vulnerabilities have been reported. With this replication results, the paper
contributes to the efforts to strengthen the empirical knowledge basis in cyber
and software security. In addition, the paper makes a contribution to the recent
discussion on the terminology about software vulnerabilities and the associated
construct validity problems that follow.
\end{abstract}

%



\received{XXX}
\received[revised]{XXX}
\received[accepted]{XXX}

\maketitle

\section{Introduction}

The paper examines what can be called a popularity hypothesis in empirical
software security research; an assumption that a software's popularity affects
the vulnerabilities reported for it. The hypothesis has been previously examined
with Java packages, the conclusion being that popularity is not a reliable
indicator of vulnerabilities reported~\cite{Siavvas18}. Also another recent
paper has evaluated the hypothesis implicitly, concluding that reported
vulnerability counts do not correlate strongly with popularity metrics for Java
packages~\cite{Sakib25}. Furthermore, a third paper has examined the hypothesis
implicitly, concluding that popular plugins for WordPress tend to gather more
vulnerabilities reported to them compared to less popular
plugins~\cite{Ruohonen19EASE}. Given that there is mixed evidence, the work
revisits the popularity hypothesis with a large-scale analysis of open source
software packages written in the PHP programming language. Two replication
assessments are presented in this regard. Both have their own replication
packages available online~\cite{Replication26}.

Within the domain of replication studies, the paper can be classified as a
dependent or a conceptual replication: although both the data and methods are
different, the paper is designed with an explicit reference to previous work in
mind~\cite{Brendel20, Fabrigar16, Gomez14, Ruohonen15COSE}, including
particularly the conclusions reached~\cite{Peels18, Shepperd18}. The
methodological difference can be justified on the grounds that a couple of the
studies replicated were both based only on correlation analysis~\cite{Siavvas18,
  Sakib25}. In contrast, the paper also uses regression analysis and
classification algorithms, which can be argued to yield stronger empirical
evidence. The recently discussed generalizability problems~\cite{Baltes22}
justify the use of different datasets. If the popularity hypothesis holds, or
does not hold, with different datasets and different methodologies, there is
either stronger or weaker evidence for it. With this point in mind, the opening
Section~\ref{sec: background} further elaborates the popularity hypothesis and
the need for a replication. The replication results using the two PHP datasets
are presented in Sections~\ref{sec: replication 1} and \ref{sec: replication
  2}. A conclusion and an accompanying discussion follow in Section~\ref{sec:
  conclusion}.

\section{Background and Related Work}\label{sec: background}

The underlying replication logic focuses on the previous work's
conclusion~\cite{Peels18}; it is important to know whether or not the hypothesis
holds with different datasets and different methods because the hypothesis is
rather fundamental in many ways. It also frequently appears in popular
discourse. The primary historical example would be Microsoft Windows. Not so
long ago, it was commonly discussed and speculated by both laypeople and
professionals that the popularity of Windows was a definite factor behind the
insecurities of that time. To quote from a famous security engineering textbook,
most ``malware writers targeted Windows rather than Mac or Linux through the
2000s and 2010s as there are simply more Windows machines to infect'', and,
furthermore, the ``model replicated itself when smartphones took over the
world''~\cite[p.~296]{Anderson08} due to the popularity of Android. Although
there is no particular reason to question these assertions, it is still worth
remarking that they are not actually backed by any empirical evidence in the
textbook. This point further motivates a need to examine the~hypothesis.

Because two replicated studies \cite{Siavvas18, Sakib25} did not support the
hypothesis but one did~\cite{Ruohonen19EASE}, the classical motivation for
replication studies to try to falsify a theory~\cite{Brendel20}, or a smaller
hypothesis, takes the form of trying to empirically affirm that the hypothesis
does in fact hold. Given a distinction between ``hard'' and ``soft''
theories~\cite{Russo19}, the popularity hypothesis is also on the side of hard
theories because it presumes a rather direct relation between a software
characteristic (popularity) and an effect (vulnerabilities reported). However,
the theorization involved is still rather soft. That is, assuming that the
hypothesis holds, the underlying explanations remain open to debate and
associated, open-ended theorization.

The starting point for a theoretical explanation given by what is now known as
(information or cyber) security economics would be
incentives~\cite{Anderson08}. That is, cyber criminals and other adversaries
have an incentive to target popular software products, packages, and projects,
whether the targeting is about finding vulnerabilities from those or explicitly
attacking them. Then: even though both commercial software vendors and open
source software projects have an incentive to develop secure code and provide
defenses, it is frequently discussed in the literature that software security
and cyber security in general are not prioritized because there are no pressing
incentives to do so~\cite{Ruohonen26EnCyCriS26}. A~slightly different
theorization path involves the famous so-called Linus law; that all bugs are
shallow when there are many eyes involved to look at code~\cite{Raymond01}. From
this alternative perspective, there are also empirical studies indicating that
widely used and hence popular open source software packages often contain more
\textit{reported} bugs and vulnerabilities than less popular
packages~\cite{Davies10, Herraiz11, Ruohonen19EASE}. These empirical studies
further reinforce the need and motivation for a replication. The italics placed
upon the word reported are also worth briefly explaining. The reason for the
italics is that empirical observations cannot generally prove that something is
secure, although these can be used to make claims about
insecurity~\cite{Herley17}. This fundamental point should be kept in mind
throughout the paper; only reported vulnerabilities are observed. By
implication, the hypothesis does not necessarily tell everything about software
security~as~such.

Finally, the paper's background is related also to the long-standing debate
about a need to align cyber security research with empirical
sciences~\cite{Herley17}. Measuring cyber security is still a grand
challenge~\cite{Meneely25}, and, at the same time, replication studies are still
being marginalized~\cite{Massacci26}. To this end, the paper also contributes to
the efforts to strengthen the empirical knowledge base in cyber security,
including with respect to validity, terminology, and theorization.

\section{The First Replication Assessment}\label{sec: replication 1}

The first replication assessment concerns a large-scale analysis of open source
software written in the PHP programming language. In what follows, the data,
methodology, and results are presented.

\subsection{Data}\label{subsec: data 1}

The dataset for the assessment was collected on 3 December 2024 from
Packagist~\cite{Packagist}, a~repository for PHP software packaged with the
Composer package manager~\cite{Composer}. In total, as many as $n = 381,993$
packages are covered in the dataset.

The dataset is composed of eight meta-data metrics provided on the Packagist's
website: (1)~\texttt{Vulnerabilities}, (2)~\texttt{Installs},
(3)~\texttt{Dependents}, (4)~\texttt{Suggesters}, (5)~\texttt{Stars},
(6)~\texttt{Watchers}, (7)~\texttt{Forks}, and (8)~\texttt{Releases}. Of these
metrics, all except \texttt{Vulnerabilities} and \texttt{Releases} can be
interpreted to proxy different dimensions of software popularity. The outlying
\texttt{Releases} metric counts the number of releases made for a given PHP
package, as reported on the Packagist's website. All metrics have a continuous
scale.

While installation amounts, \texttt{Installs}, provide a rather straightforward
and widely used metric for software popularity~\cite{Herraiz11, Paramitha13,
  Ruohonen19EASE, Kula18, Qiu18}, the \texttt{Dependents} metric conveys a
package's popularity in terms of incoming edges in a dependency network between
packages~\cite{Kula18}. With this metric, a package is seen as popular when many
other packages depend on it. Both installation (or download) counts and
dependencies have also been used to help security-related decision-making about
funding applications for open source software
projects~\cite{Ruohonen26EnCyCriS26}. Although documentation is lacking,
\texttt{Suggesters} generally proxy different ``recommended'' dependencies; for
each package, Packagist recommends further packages in addition to a package's
conventional dependencies, as captured by \texttt{Dependents}, without which the
package does not install or function properly.

The \texttt{Stars}, \texttt{Watchers}, and \texttt{Forks} metrics are explicitly
tied to functionality provided by GitHub on which almost all of the actual
development of the PHP packages occurs today. In general, with ``stars''
developers can flag packages and projects they perceive as interesting or
relevant. By adding oneself as a ``watcher'' to a project, one can get
information about changes in the project. The \texttt{Forks} metric, in turn,
refers to the number of times a given project was forked, as has been common in
the open source world throughout the decades~\cite{Gencer12}. According to
practitioner surveys, all three metrics have been seen as relevant for proxying
popularity~\cite{Borges18}. Finally, the dependent metric is
\texttt{Vulnerabilities}, which counts and records vulnerabilities reported for
a package. Many (but not all) of the reported vulnerabilities tracked on
Packagist are identified with Common Vulnerabilities and Exposures~(CVEs), while
the associated security advisories typically point toward GitHub. In this
regard, it can be noted that no additional data were retrieved to validate the
reported CVE-referenced vulnerability information provided on Packagist. Nor
were GitHub or other hosting sites queried for retrieving further software
development~data.

\subsection{Methods}\label{subsec: methods 1}

Imbalance is a highly typical and difficult problem in computational cyber
security applications~\cite{Hague25, Singh22, Sayegh25, Wheelus18}. As soon
described in Subsection~\ref{subsec: results 1 descriptive statistics}, also the
Packagist dataset is highly imbalanced, meaning that no vulnerabilities have
been reported for most of the packages. The imbalance also justifies the
methodology.

\subsubsection{Descriptive Statistics}

One of the papers replicated \cite{Siavvas18} used correlation analysis to draw
the conclusion that the hypothesis would not hold. However, correlation analysis
is known to be problematic with long-tailed data~\cite{Barber20, Bishara15,
  Gil17}. Therefore, Pearson's correlation coefficients are presented merely to
observe whether an incorrect conclusion would be drawn by relying solely on
them. They also provide information for forming a sum~variable:
\begin{align}\label{eq: sum variable}
  \texttt{SumVariable}_i &= (\texttt{Installs}_i + \texttt{Dependents}_i +
  \texttt{Suggesters}_i \\ \notag &+ \texttt{Stars}_i + \texttt{Watchers}_i +
  \texttt{Forks}_i)~/~6 .
\end{align}

If all six metrics measure software popularity, as was assumed earlier in
Subsection~\ref{subsec: data 1}, Cronbach's standardized
$\alpha$-coefficient~\cite{Cronbach51} should also attain a relatively high
value. For the popularity hypothesis to hold, an auxiliary hypothesis
($\textmd{H}_{11}$) can also be presented: the arithmetic means are different
for each metric appearing in~\eqref{eq: sum variable} when separated into two
groups according to the following truncated $\texttt{Vulnerabilities}^b$ metric:
\begin{equation}\label{eq: binary valued vulnerabilities}
\texttt{Vulnerabilities}^b_i =
\begin{cases}
0\quad\textmd{if}\quad\texttt{Vulnerabilities}_i = 0 , \\
1\quad\textmd{if}\quad\texttt{Vulnerabilities}_i > 0 .
\end{cases}
\end{equation}
The auxiliary hypothesis $\textmd{H}_{11}$ is tested with the $t$-test, using a
correction~\cite{Welch47} for unequal variances between the groups. The
correction is necessary because variance is much higher in the
$\texttt{Vulnerabilities}_i = 0$~group.

\subsubsection{Regression Analysis}\label{subsec: regression analysis 1}

Three regression models are estimated:
\begin{align}\label{eq: reg 1}
\ln(\texttt{Vulnerabilities}_i + 1) = \alpha_1 &+ \beta_{11}\ln(\texttt{Releases}_i + 1) \\ \notag &+ \epsilon_{1i}, \\
\ln(\texttt{Vulnerabilities}_i + 1) = \alpha_2 &+ \beta_{21}\ln(\texttt{SumVariable}_i + 1) \\ \notag &+ \beta_{22}\ln(\texttt{Releases}_i + 1) + \epsilon_{2i} ,
\\
\ln(\texttt{Vulnerabilities}_i + 1) = \alpha_3 &+ \beta_{31}\ln(\texttt{Installs}_i + 1) \\ \notag &+ \beta_{32}\ln(\texttt{Dependents}_i + 1) \\ \notag &+ \beta_{33}\ln(\texttt{Suggesters}_i + 1) \\ \notag &+ \beta_{34}\ln(\texttt{Stars}_i + 1) \\ \notag &+ \beta_{35}\ln(\texttt{Watchers}_i + 1) \\ \notag &+ \beta_{36}\ln(\texttt{Forks}_i + 1) \\ \notag &+ \beta_{37}\ln(\texttt{Releases}_i + 1) + \epsilon_{3i} .
\end{align}

The three models are estimated with the conventional ordinary least squares
(OLS). However, there is a long-standing debate over whether the
log-transformation, $\ln(\texttt{Vulnerabilities}_i + 1)$, can be used for count
data. The general conclusion seems to be that OLS with the transformation can be
used for testing statistical significance~\text{\cite{Ives15, StPierre17}}, but
the actual coefficient estimates can be biased~\cite{StPierre17}. Therefore, the
interest is to only observe whether the overall statistical performance
increases from the first model to the second and third models. If the popularity
hypothesis holds, an auxiliary hypothesis~($\textmd{H}_{12}$) that
$\hat{\beta}_{21} > 0$, $\hat{\beta}_{31} > 0$, $\hat{\beta}_{32} > 0$,
$\hat{\beta}_{33} > 0$, $\hat{\beta}_{34} > 0$, $\hat{\beta}_{35} > 0$, and
$\hat{\beta}_{36} > 0$ should also hold in addition to another auxiliary
hypothesis~($\textmd{H}_{13}$) that all estimated coefficients in the set
$\lbrace \hat{\beta}_{21}, \hat{\beta}_{31}, \hat{\beta}_{32}, \hat{\beta}_{33},
\hat{\beta}_{34}, \hat{\beta}_{35}, \hat{\beta}_{36} \rbrace$ are statistically
significant at a 95\% confidence~level.

\subsubsection{Classification}

Machine learning classification is further used with the binary-valued
$\texttt{Vulnerabilities}^b$ metric. Three conventional algorithms are used:
na\"ive Bayes~\cite{Majka24}, (boosted) logistic regression~\cite{Tuszynski24},
and random forest~\cite{Liaw02}. All are well-known and thus require no
particular elaboration. Instead, the implementations used are worth remarking;
these are all R packages implemented to work with the \texttt{caret}
package~\cite{Kuhn08}. Default parameters were used for all algorithms. The only
exception is the random forest classifier implementation for which the maximum
number of trees had to be restricted to one hundred due to memory
constraints. For each algorithm, three distinct sets of explanatory metrics are
used analogously to the regression models in \eqref{eq: reg 1}. In other words,
the first set contains only \texttt{Releases}, the second set
\texttt{SumVariable} and \texttt{Releases}, and the third set the seven metrics.

For each algorithm and for each set, four solutions are used for dealing with
the class imbalance: (1)~oversampling, (2)~undersampling,
(3)~SMOTE~~\cite{Chawla02, Torgo13}, and ROSE~\cite{Lunardon14, Menardi14}. Of
these sampling solutions, the details of which can be found from the literature
referenced, particularly SMOTE is commonly seen as a today's \textit{de~facto}
solution for handling imbalanced data in the classification
context~\cite{Fernandez18}. Regardless, it should be emphasized that the goal is
\textit{not} about maximizing performance but about testing of a hypothesis;
therefore, the interest is to observe whether the results are consistent across
the three classification and four balancing~algorithms.

As for practical computation, the $\ln(x + 1)$ transformation is applied to all
explanatory metrics. A $10$-fold cross-validation is used for training. Testing
is done with a random sample containing 25\% of the dataset. Due to the
logarithm transformation, neither scaling nor centering are used for the
explanatory metrics. Finally, it should be mentioned that the class imbalance
also affects metrics for evaluating classification performance. Among other
things~\cite{Menardi14}, the traditional accuracy metric is biased and
misleading in this context~\cite{Flores25}. Therefore, three alternative metrics
are used.

First, a simple balanced accuracy (BA) has been used in existing
work~\cite{Brodersen10}. It is given by $\textmd{BA} = (\textmd{TPR} +
\textmd{TNR})~/~2$, where the true positive rate (TPR) and the true negative
rate (TNR)~are $\textmd{TPR} = \textmd{TP}~/~(\textmd{TP} + \textmd{FN})$ and
$\textmd{TNR} = \textmd{TN}~/~(\textmd{TN} + \textmd{FP})$, where, in turn, TP
refers to true positives, FN to false negatives, TN to true negatives, and FP to
false positives. The TPR metric is also known as a recall or a sensitivity and
TNR as a specificity. Second, a G-mean metric has also been used in the
imbalanced classification context~\cite{Kaleeswari23, Susan20}. It is defined as
$\textmd{G-mean} = \sqrt{\textmd{TPR} \times \textmd{TNR}}$. The BA and G-mean
metrics vary in the unit interval. Higher values are better. Third, the
conventional mean squared error (MSE), which is sometimes also known as the
Brier score~\cite{Flores25, Kaleeswari23}, has further been used in the
imbalanced classification context. It is given by the usual formula:
$\textmd{MSE} = (1/n)\sum^n_{i=1} (\widehat{p}_i - p_i)^2$, where
$\widehat{p}_i$ is an estimated probability for a package's release history
having seen at least one reported vulnerability, while $p_i$ is the actual
probability; $p_i = 1$ in case at least one vulnerability has been reported for
the $i$:th package. Unlike with the BA and G-mean metrics, lower values are
better.

\subsection{Results}\label{sec: results 1}

\subsubsection{Descriptive Statistics}\label{subsec: results 1 descriptive statistics}

\begin{figure*}[th!b]
\centering
\includegraphics[width=16cm, height=5cm]{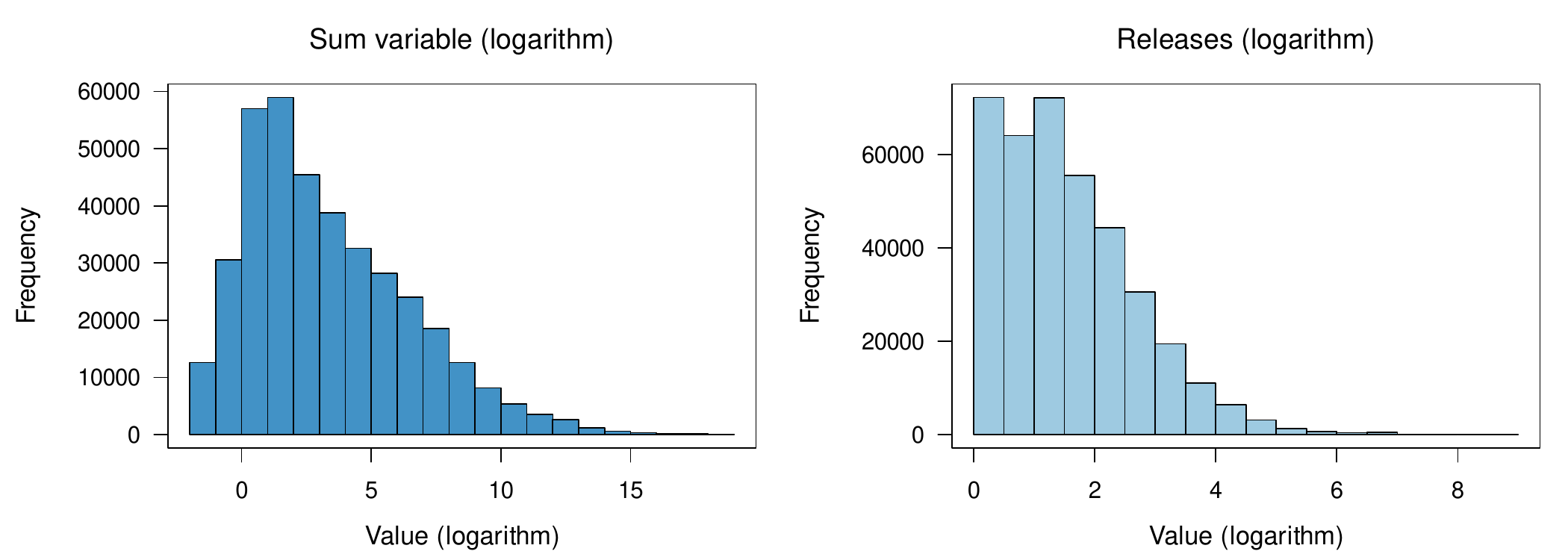}
\caption{Distributions of the Popularity Sum Variable and \textit{Releases}}
\label{fig: hist ind}
%
\vspace{35pt}
%
\centering
\includegraphics[width=14cm, height=9cm]{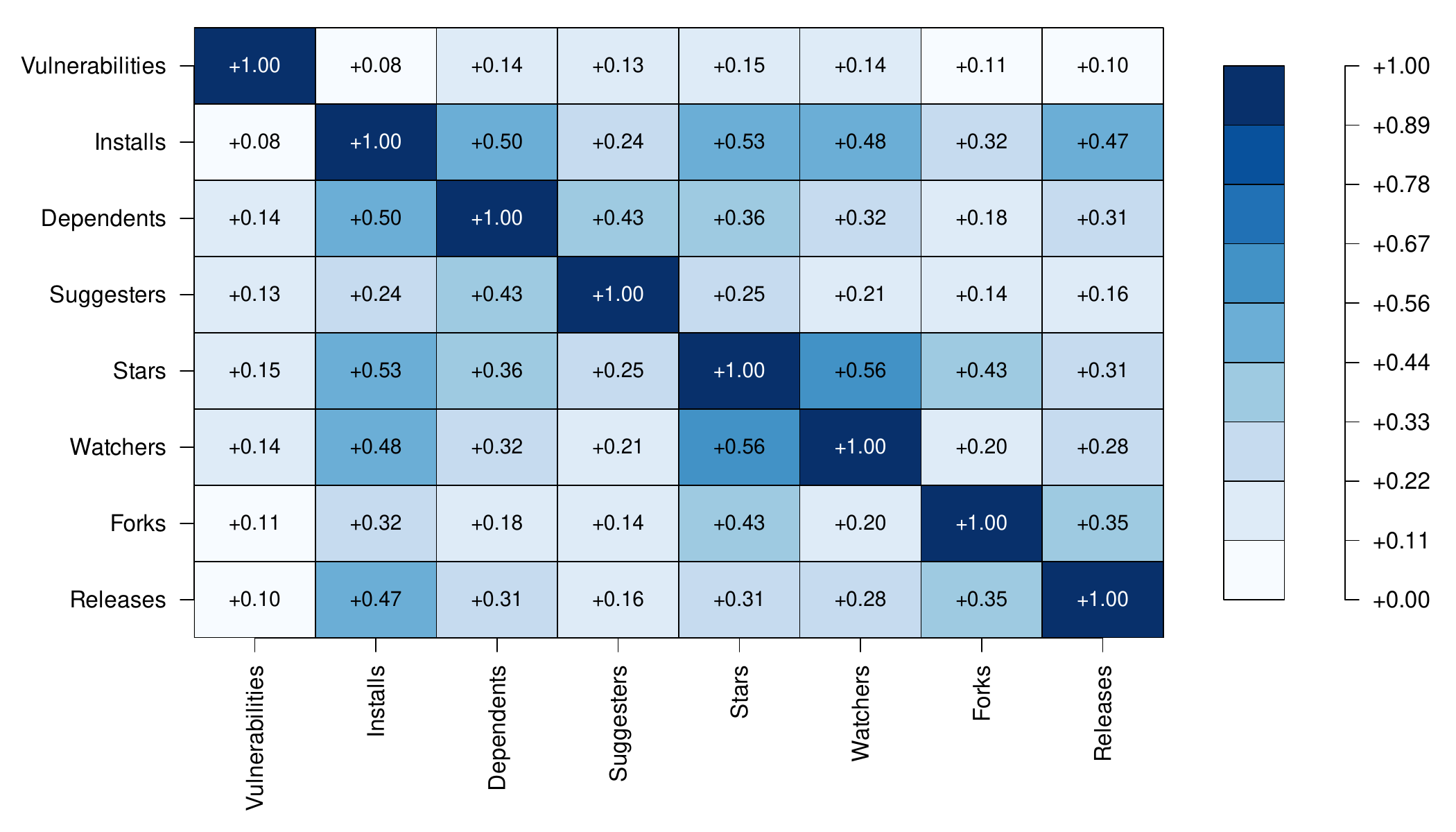}
\caption{Correlations in the First Dataset}
\label{fig: correlations}
\end{figure*}

The presentation of the results can be started by reiterating the earlier
methodological points. Thus, the extreme imbalance is best illustrated with
plain numbers: of the over $381$ thousand PHP packages, only as few as $777$, or
about $0.2\%$, have seen one or more reported vulnerabilities over their release
histories. When taking a look at Table~\ref{tab: vulnerabilities}, it can be
also concluded that a single vulnerability has been reported for the
majority~(57\%) of the $777$ packages with a record of
reported~vulnerabilities. However---and as has been previously observed with PHP
packages~\cite{Ruohonen19EASE}, the tail is relatively long; there are outlying
packages that have seen nine or more reported vulnerabilities. A~similar
observation applies to the remaining metrics; they all have long tails. As can
be further seen from Fig.~\ref{fig: hist ind}, also the popularity sum variable
and the \texttt{Releases} metric have a long-tailed distribution. This
observation too is familiar from existing studies~\cite{Sakib25,
  Ruohonen18IWESEP}. Therefore, also the $\ln(x + 1)$ transformation
is~justified.

\begin{table*}[th!b]
\centering
\caption{Reported Vulnerabilities Across the PHP Packages}
\label{tab: vulnerabilities}
\begin{tabular}{llrcrcrcrcrcrcrcrcr}
\toprule
&& \multicolumn{17}{c}{Number of vulnerabilities reported} \\
\cmidrule{3-19}
&& 0 && 1 && 2 && 3 && 4 && 5 && 6 && 7 && $\geq$ 8 \\
\hline
Frequency && $381,216$ && $433$ && $114$ && $51$ && $27$ && $23$ && $15$ && $7$ && $107$ \\
Share (\%) && $99.797$ && $0.113$ && $0.030$ && $0.013$ && $0.007$ && $0.006$ && $0.004$ && $0.002$ && $0.028$ \\
\bottomrule
\end{tabular}
\vspace{35pt}
\end{table*}

\begin{figure*}[th!b]
\centering
\includegraphics[width=16cm, height=3.5cm]{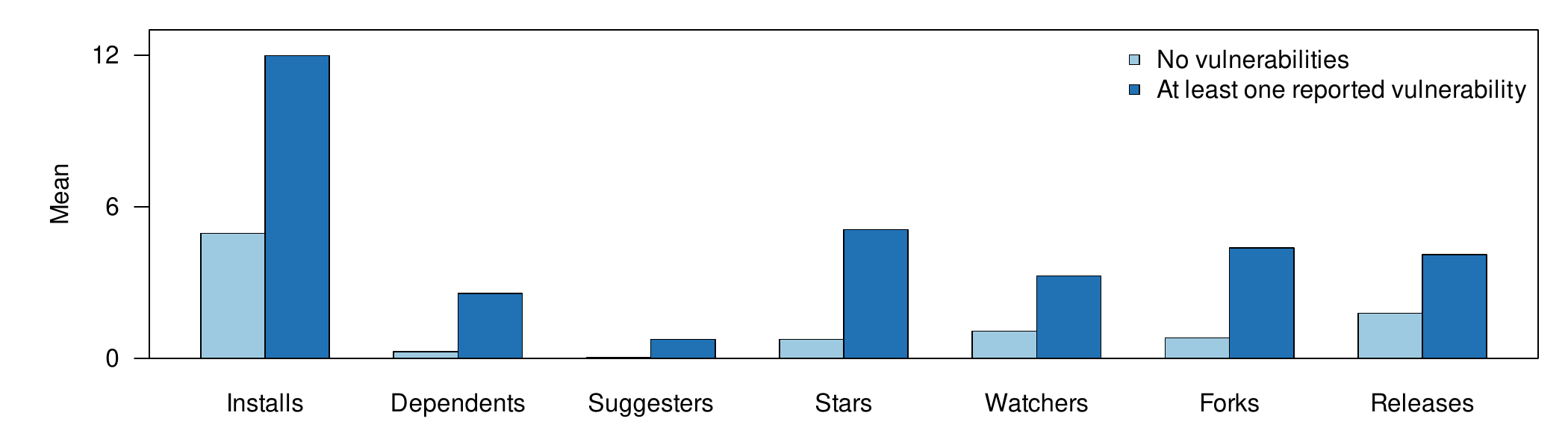}
\caption{Means Across Vulnerability Groups (all $t$-tests statistically significant at \text{$p < 0.001$})}
\label{fig: tt}
%
\vspace{15pt}
%
\centering
\includegraphics[width=16cm, height=5cm]{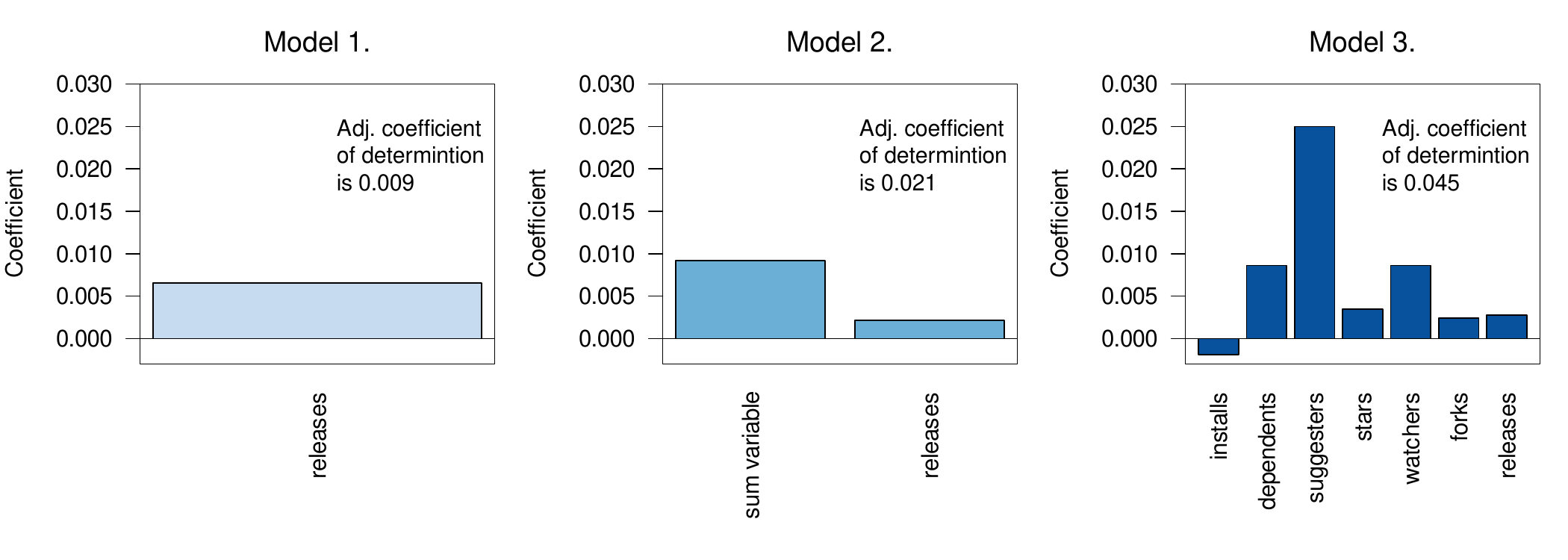}
\caption{Regression Results (OLS, full sample)}
\label{fig: ols}
\vspace{15pt}
\end{figure*}
\begin{table*}[th!b]
\centering
\caption{Classification Results$^1$}
\label{tab: classification}
\begin{threeparttable}
\begin{tabular}{lrrrcrrrcrrr}
\toprule
& \multicolumn{3}{c}{Na\"ive Bayes} && \multicolumn{3}{c}{Boosted LR} && \multicolumn{3}{c}{Random Forest} \\
\cmidrule{2-4}\cmidrule{6-8}\cmidrule{10-12}
& \multicolumn{3}{c}{Model} && \multicolumn{3}{c}{Model} && \multicolumn{3}{c}{Model} \\
\cmidrule{2-4}\cmidrule{6-8}\cmidrule{10-12}
& \qquad 1. & 2. & 3. && 1. & 2. & 3. && 1. & 2. & 3. \\
\hline
BA \\
\cmidrule{1-1}
Oversampling & 0.771 & 0.726 & 0.749 && 0.731 & 0.754 & 0.805 && 0.505 & 0.543 & 0.508 \\
Downsampling & 0.771 & 0.736 & 0.803 && 0.793 & 0.633 & 0.761 && 0.704 & 0.689 & 0.772 \\
SMOTE & 0.771 & 0.726 & \e{0.820} && 0.731 & 0.759 & 0.816 && 0.524 & 0.599 & 0.525 \\
ROSE & 0.771 & 0.726 & 0.809 && 0.744 & 0.734 & \e{0.840} && 0.625 & 0.696 & \e{0.825} \\
\cmidrule{2-4}
G-mean \\
\cmidrule{1-1}
Oversampling & 0.759 & 0.678 & 0.747 && 0.695 & 0.722 & 0.796 && 0.484 & 0.319 & 0.125 \\
Downsampling & 0.759 & 0.693 & 0.803 && 0.788 & 0.522 & 0.732 && 0.694 & 0.622 & 0.746 \\
SMOTE & 0.759 & 0.678 & \e{0.819} && 0.695 & 0.727 & 0.814 && 0.405 & 0.525 & 0.231 \\
ROSE & 0.759 & 0.678 & 0.807 && 0.717 & 0.692 & \e{0.840} && 0.586 & 0.637 & \e{0.824} \\
\cmidrule{2-4}
MSE & \\
\cmidrule{1-1}
Oversampling & 0.283 & 0.463 & \e{0.178} && 0.492 & 0.383 & 0.260 && 0.353 & 0.018 & \e{0.007} \\
Downsampling & 0.292 & 0.442 & 0.206 && 0.492 & 0.383 & 0.260 && 0.327 & 0.496 & 0.314 \\
SMOTE & 0.284 & 0.463 & 0.199 && 0.492 & 0.366 & 0.914 && 0.145 & 0.094 & 0.008 \\
ROSE & 0.278 & 0.460 & 0.219 && 0.455 & 0.425 & \e{0.118} && 0.464 & 0.464 & 0.104 \\
\bottomrule
\end{tabular}
\begin{tablenotes}
\begin{scriptsize}
\vspace{3pt}
\item{$^1$~The best values are colored in each of the three rowwise panels.}
\end{scriptsize}
\end{tablenotes}
\end{threeparttable}
\end{table*}

Regarding multicollinearity, Fig.~\ref{fig: correlations} displays Pearson's
correlation coefficients across all metrics in the dataset by using the
logarithm transformation. As can be seen, all coefficients have positive signs
and some of these are relatively large in their magnitudes. These correlations
are nothing surprising as such. Particularly when operating with large
behavioral \text{datasets---a} domain to which the dataset examined can also be
seen to belong, everything tends to be correlated with
everything~\cite{Meeh90}. In any case, \texttt{Stars}, \texttt{Watchers}, and
\texttt{Installs} are moderately or even strongly correlated with each other on
one hand and \texttt{Installs} and \texttt{Releases} on the other, to use the
thresholds and adjectives for these from one of the replicated
studies~\cite{Siavvas18}. These correlations align with the Cronbach's
standardized $\alpha$-coefficient value of $0.64$ for the \texttt{SumVariable}
defined in \eqref{eq: sum variable}. Given the maximum of one, the value is not
particularly high but described in the literature with such adjectives as
acceptable, satisfactory, and sufficient have been used to describe comparable
values in existing research~\cite{Taber17}. Thus, all in all, the six metrics
can indeed be interpreted to measure the same underlying software popularity
concept, as was also presumed.

The observations are in contrast with one replicated study, which concluded that
reported vulnerability counts ``show almost no relationship with metrics like
stars or forks'' \cite[p.~4]{Sakib25}. Furthermore, these correlations may
affect the classifications because particularly the na\"ive Bayes and logistic
regression (LR) rely on an independence assumption. Though, even na\"ive Bayes
seems to still perform well under
multicollinearity~\cite{Araveeporn24}. Therefore, it is more relevant to
continue by pointing out the only modest magnitudes between
\texttt{Vulnerabilities} and the rest of the metrics. This observation might be
taken as a prior expectation that the hypothesis may not hold.

However, the $t$-tests visualized in Fig.~\ref{fig: tt} tell a different
story. As can be seen, all means are different in the two groups separated by
the truncated \texttt{Vulnerabilities}$^b$ metric. In other words, both popular
packages and packages with long release histories seem to have witnessed more
reported vulnerabilities over the years. The auxiliary hypothesis
$\textmd{H}_{11}$ holds. This conclusion provides a good motivation to continue
to the regression~results.

\subsubsection{Regression Analysis}

Regression results can shed a little more light on the earlier correlation
results in Fig.~\ref{fig: correlations}. Thus, coefficients from three OLS
regressions for the three models specified in \eqref{eq: reg 1} are shown in
Fig.~\ref{fig: ols}. The logarithm transformation is again used for all
metrics. The auxiliary hypothesis $\textmd{H}_{13}$ holds; all coefficients are
statistically significant at a 99\% confidence level, which is hardly surprising
due to the sample size~\cite{Bakan66}. As could be furthermore expected, the
performance increases the more there are metrics. The full model yields an
adjusted $R^2 = 0.045$, meaning that roughly about five percent of the total
variance is explained by the seven metrics. When keeping in mind that there are
only seven hundred seventy-seven observations for which \texttt{Vulnerabilities}
attains a value larger than zero, the performance is not necessarily that
bad. Of the coefficients for the individual metrics, the one for
\texttt{Suggesters} stands out. Interestingly, the coefficient for
\texttt{Installs} has a negative sign, which seems to contradict the earlier
results in Fig.~\ref{fig: tt}. Therefore, as specified in
Subsection~\ref{subsec: regression analysis 1}, the auxiliary hypothesis
$\textmd{H}_{12}$ does not hold. All in all, nevertheless, the evidence is again
on the side of supporting the popularity hypothesis. The magnitudes and signs of
the coefficients are also likely linked to the correlations between the metrics.

\subsubsection{Classification}

The classification results are summarized in Table~\ref{tab: classification} for
the three classification algorithms, three models, and four balancing
algorithms. To recall: the first model is the barebone one with just the
\texttt{Releases} metric, the second model adds the sum variable of the
popularity metrics, and the third model uses all seven metrics
individually. With this remark in mind, the discussion of the classification
results can be started by noting that the balancing algorithms improve the
performance significantly. Although not reported, plain unbalanced
classifications yield much worse performance, regardless of the three
classifiers and the three models. Regarding the balancing algorithms, it seems
that the ROSE algorithm is better than the rest, although SMOTE does well with
the Na\"ive Bayes classifier, and plain oversampling yields the best performance
in one outlying case. As for the classifiers, the boosted logistic regression
outperforms the Na\"ive Bayes and random forest classifiers according to the BA
and G-mean metrics. According to these two performance metrics, the highest
value reached is~$0.840$, which can be interpreted as a decent value in the
context of extreme imbalance and limited explanatory information used. Most
importantly, the best values colored in each table refer to the third model. In
all cases these are much better than with the first model. In conclusion: the
popularity hypothesis can be taken to hold.

\section{The Second Replication Assessment}\label{sec: replication 2}

The second replication assessment concerns an empirical analysis of third-party
open source software components written for the popular WordPress web
framework. Henceforth, the term component refers to plugins and themes provided
for WordPress. In what follows, the data, methodology, and results presented
consecutively.

\subsection{Data}\label{subsec: data 2}

The dataset for the second replication assessment was collected from two sources
on 6 December 2025. The first is the vulnerability data feed for WordPress and
its components provided by Wordfence~\cite{Wordfence25}. Excluding the core
WordPress itself, each plugin and theme present in the feed at the time of data
collection was queried from the WordPress' homepage~\cite{WordPress25a} to
retrieve popularity data and associated meta-data. Only those components that
were present in both sources are included in the dataset. In addition, the
dataset was reduced by excluding components with missing values. These missing
values were present for installation counts, WordPress versions supported by the
components, and last updating dates. The statistical analysis operates with
$6,655$ components.

Regarding software popularity, the WordPress site provides two metrics. The
first is approximate installation amounts. These are based on telemetry
data. Whenever a WordPress deployment uses a third-party component, it will ping
a plugin database to see whether updates are available. The second popularity
metric is based on the star-ratings provided on the WordPress homepage. As
usual, these are used by users to vouch for some plugins and themes or downvote
others. When compared to the installation counts, the star-ratings are
user-dependent and can hence be unreliable. Recent research has also provided
evidence about gaming of star-ratings on GitHub~\cite{He26}. For these reasons,
the installation counts serve as the primary material for the assessment. The
assessment is carried out with three vulnerability count metrics, which are
subsequently described alongside the statistical regression methodology.

\subsection{Methods}\label{subsec: methods 2}

Also the second replication assessment uses regression analysis. After
elaborating the methodology in this regard, the operationalization of the
explanatory metrics is discussed in relation to the regression specification used.

\subsubsection{Regression Analysis}

The following specification is used:
\begin{align}\label{eq: wp ols}
  y_i = \alpha
  +& \beta_1(\texttt{Installs:~[101, 1000]})_i \\ \notag
  +& \beta_2(\texttt{Installs:~[1001, 10000]})_i \\ \notag
  +& \beta_3(\texttt{Installs:~[10001, 100000]})_i \\ \notag
  +& \beta_4(\texttt{Installs:~[100001, $\infty$]})_i \\ \notag
  +& \beta_5(\texttt{WordPress:~4 series})_i \\ \notag
  +& \beta_6(\texttt{WordPress:~5 series})_i \\ \notag
  +& \beta_7(\texttt{WordPress:~6-7 series})_i \\ \notag
  +& \beta_8(\texttt{Themes})_i \\ \notag
  +& \beta_9(\texttt{Removed})_i \\ \notag
  +& \beta_{10}(\texttt{TopReporters})_i \\ \notag
  +& \beta_{11}(\texttt{Reporters})_i \\ \notag
  +& \beta_{12}(\texttt{LastUpdated})_i \\ \notag
  +& \beta_{13}(\texttt{Tags})_i \\ \notag
  +& \beta_{14}(\texttt{CVSS})_i \\ \notag
  +& \beta_{15}(\texttt{Stars: 4+5})_i \\ \notag
  +& \beta_{16}(\texttt{Stars: 1+2+3})_i \\ \notag
  +& \beta_{17}(\texttt{Outlier})_i
  + \epsilon_i ,
\end{align}
where $y_i$ is either (a)~counts of all vulnerabilities reported, (b)~counts of
reported and patched vulnerabilities, or (c)~counts of reported but unpatched
vulnerabilities for a WordPress component $i$ among the $n = 6,655$ components
observed, $\alpha$ is again constant, $\beta_1, \ldots, \beta_{17}$ are
regression coefficients, and, finally, $\epsilon_i$ is a residual term.

The regression specification is estimated with a so-called quasi-Poisson
regression, as also done in one of the papers
replicated~\cite{Ruohonen19EASE}. It is similar to the negative binomial
regression with a difference that a linear correction instead of a quadratic one
is done to account for overdispersion~\cite[p.~403]{DunnSmyth18}. Further
details about the estimation are provided when presenting the results in
Subsection~\ref{subsec: regression analysis 2}.

\subsubsection{Hypotheses}

Of the regression coefficients, a particular interest is on the four
coefficients $\beta_1, \ldots, \beta_4$, which represent the statistical effects
compared to the least popular category for WordPress components having only up
to a hundred installations at maximum. For the popularity hypothesis to hold,
the following three auxiliary hypotheses should hold:
\begin{enumerate}[label=H$_{2\arabic*}$:]
\itemsep 3pt
\item{the estimated coefficients $\hat{\beta}_1, \ldots, \hat{\beta}_4$ are
  statistically significant at the conventional 95\% confidence level;}
\item{$\hat{\beta}_1 < \hat{\beta}_2 <   \hat{\beta}_3 < \hat{\beta}_4$ for the counts of all vulnerabilities and patched vulnerabilities reported; and}
\item{$\hat{\beta}_1 > \hat{\beta}_2 > \hat{\beta}_3 > \hat{\beta}_4$ for the
  counts of reported but unpatched vulnerabilities at the time of the
  data~collection.}
\end{enumerate}
\vspace{3pt}

If $\textmd{H}_{22}$ holds, the monotonically increasing effects signify that
counts of all vulnerabilities reported and patched vulnerabilities reported
increase with increasing installation counts; the more there are online
installations running with given components, the more there are vulnerabilities
reported for the components. If the hypothesis $\textmd{H}_{23}$ further holds,
less popular WordPress components have seen more reported yet unpatched
vulnerabilities. In addition, a fourth auxiliary hypothesis ($\textmd{H}_{24}$)
can be presented: $\hat{\beta}_{15}$ is statistically significant at the 95\%
confidence level and $\hat{\beta}_{15} > \hat{\beta}_{16}$. If this hypothesis
holds too, higher star-ratings by users would entail higher vulnerability
counts and the effect would be logically consistent when compared to lower
star-ratings by users.

\subsubsection{Operationalization}\label{subsec: wp operationalization}

The operationalization of the metrics in the regression specification can be
elaborated in the order they appear in \eqref{eq: wp ols}. Thus, to begin with,
the installation counts were re-categorized from the more fine-grained
installation groups used in the online source. The re-categorized groups are
almost the same as in previous research~\cite{Ruohonen19EASE}. The reference
group for the four estimated $\hat{\beta}_1$, $\hat{\beta}_2$, $\hat{\beta}_3$,
and $\hat{\beta}_4$ is \texttt{(Installs:~[0, 100])}.

Analogously, the reference group for the three subsequent coefficients is a
dummy variable recording support for the first, second, and third major
WordPress release series or any releases thereafter. As WordPress follows a
``major.minor'' semantic versioning scheme, the major release series were
identified from the leading numbers provided. For instance, a meta-data string
``\texttt{5.8 or higher}'' for a plugin~\cite{WordPress25b} implies that the
\texttt{(WordPress: 5 series)} dummy variable takes a value of one for this
particular plugin. Also \texttt{Themes} and \texttt{Removed} are dummy variables
denoting whether a component is a theme (instead of a plugin) and whether a
component has been removed from the WordPress' homepage for maintenance or other
problems. Although documentation is lacking, the presence of unpatched
vulnerabilities for a relatively long time would presumably be something that
prompts a removal. Furthermore, two additional dummy variables are used; the
first, \texttt{TopReporters}, records whether a particularly productive
vulnerability reporter has reported vulnerabilities to a given plugin. The
second is a statistical control for extreme outliers. As these two dummy
variables are included on statistical rather than theoretical or contextual
grounds, they are elaborated later during presenting descriptive statistics.

All remaining metrics in the regression specification have a continuous
scale. The \texttt{LastUpdated}, which has been used also
previously~\cite{Ruohonen19EASE}, counts the days since the last update to a
given component. Then, \texttt{Tags} counts the number of tags present for a
given component. For instance, a value five is present for the tags for the
example plugin referenced earlier~\cite{WordPress25b}. The subsequent
\texttt{CVSS} metric approximates the average severity of all vulnerabilities
reported for a component with the Common Vulnerability Scoring System (CVSS). It
is operationalized as the median of CVSS (v.~4) base scores of the
vulnerabilities reported for a given component. The WordPress site provides also
the online star-rating functionality. Due to multicollinearity issues, the
counts of four-star and five-star ratings were summed into a single
variable. Likewise, $(\texttt{Stars: 1+2+3})$ records a sum of one-star,
two-star, and three-star counts for a component. Finally, it is should be
emphasized that the metrics used are not sufficient for seeking maximal
statistical performance in general. Instead, the interest is again to examine
the popularity hypothesis, including with respect to $\textmd{H}_{21}$,
$\textmd{H}_{22}$, $\textmd{H}_{23}$, and $\textmd{H}_{24}$.

\subsection{Results}

\subsubsection{Descriptive Statistics}\label{subsec: desc stats 2}

\begin{figure*}[th!b]
\centering
\includegraphics[width=16cm, height=4.5cm]{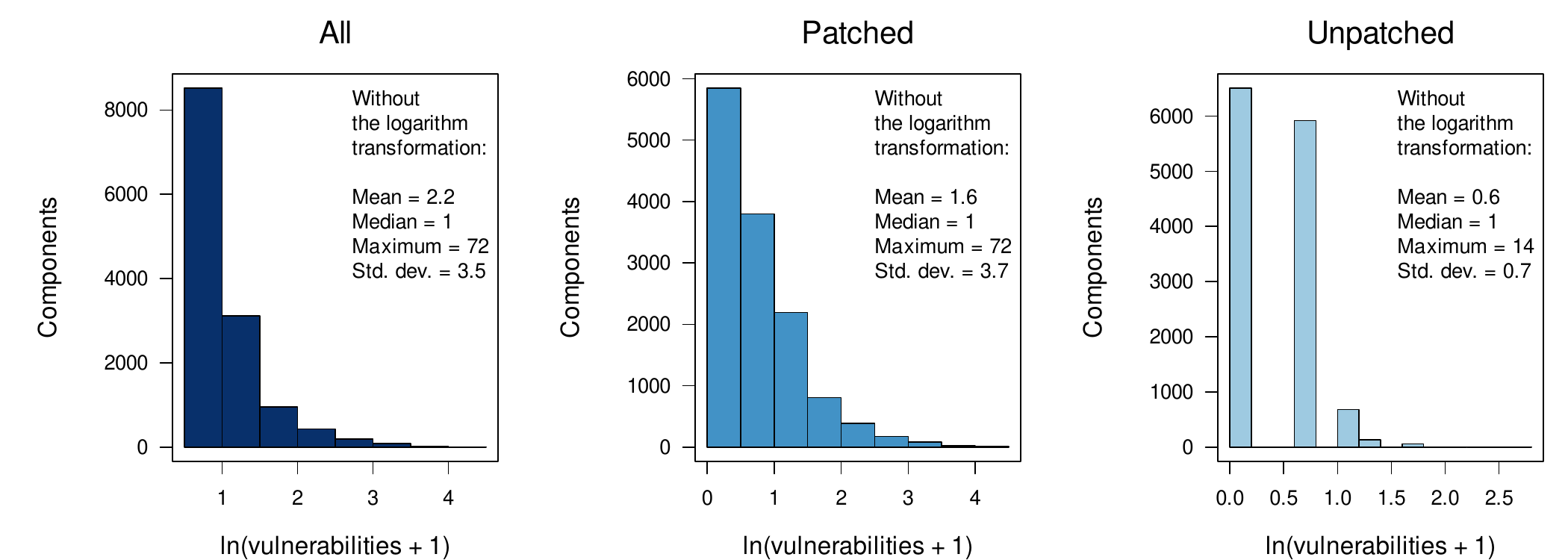}
\caption{All Vulnerabilities, Patched Vulnerabilities, and Unpatched Vulnerabilities Reported for the WordPress Components}
\label{fig: wp vulns}
%
\vspace{15pt}
%
\centering
\includegraphics[width=16cm, height=4.4cm]{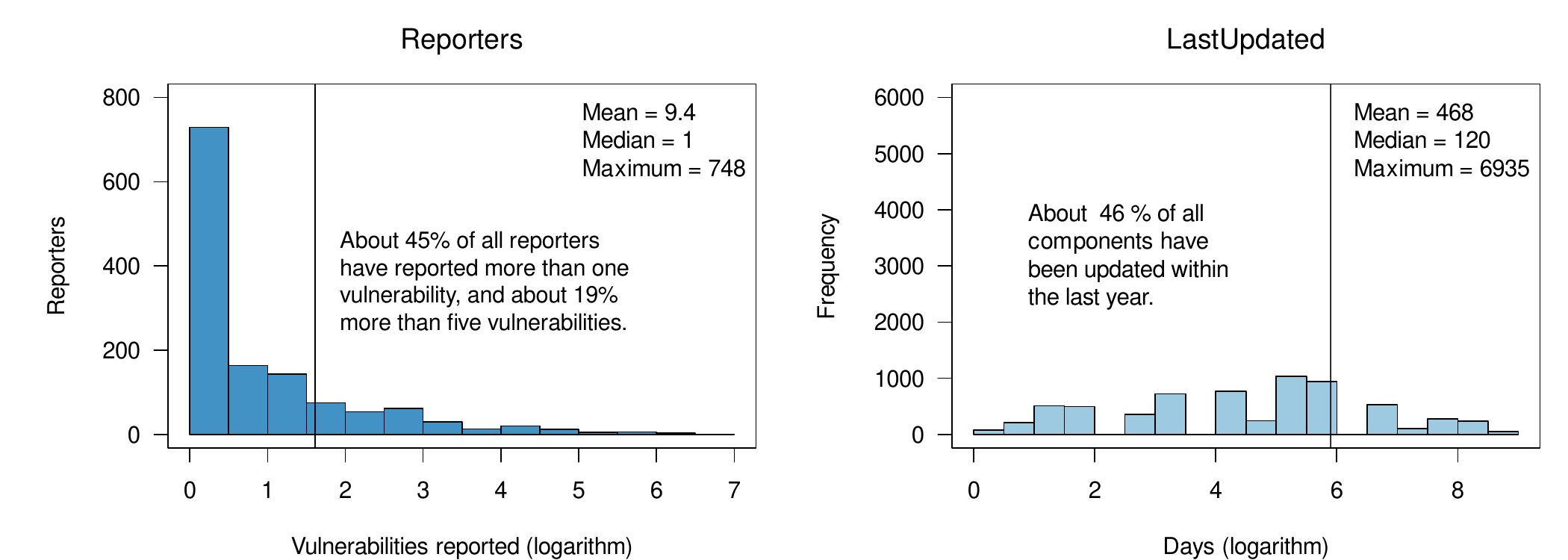}
\caption{Reporters of Vulnerabilities and Updating Delays for the WordPress Components}
\label{fig: wp resup}
%
\vspace{15pt}
%
\centering
\includegraphics[width=16cm, height=4.4cm]{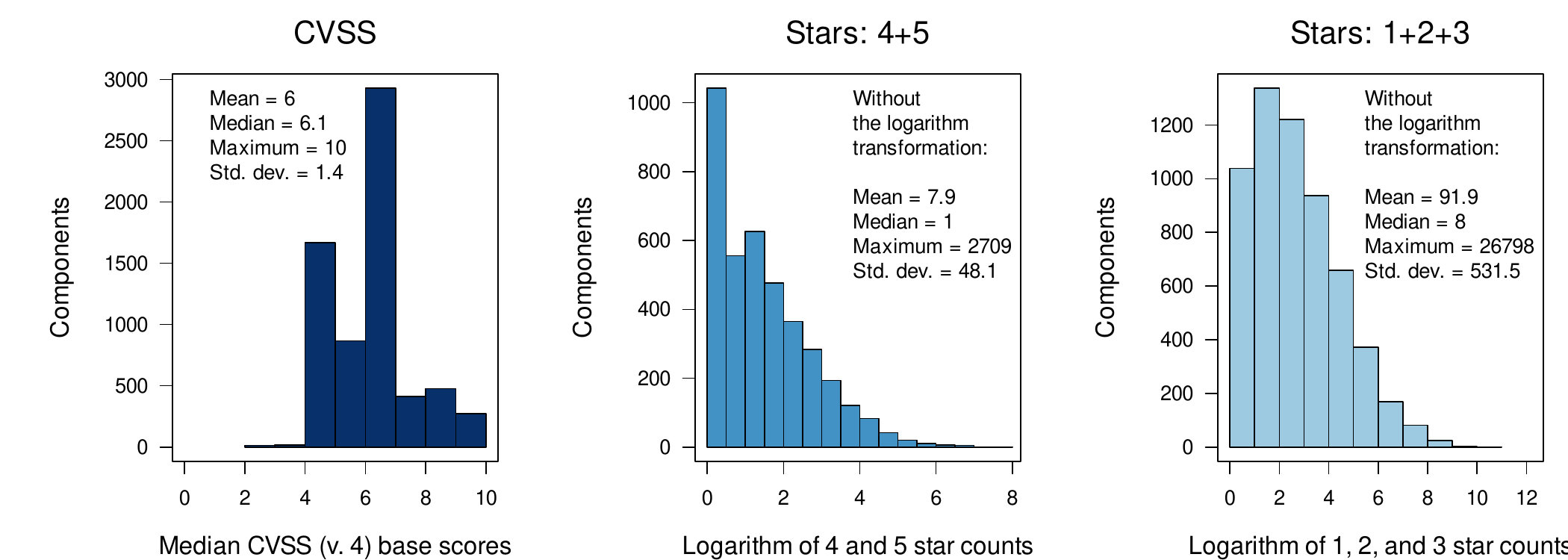}
\caption{Median CVSS (v.~4) Base Scores and Star Rankings for the WordPress Components}
\label{fig: wp cvssstars}
\end{figure*}

The three vulnerability counts all have extremely long-tailed distributions. As
can be seen from Fig.~\ref{fig: wp vulns}, even after the $\ln(x + 1)$
transformation, the empirical distributions still resemble long-tailed
probability distributions. Regarding all vulnerabilities reported, only a single
vulnerability has been reported for about 64\% of the components
observed. However, the maximum is as large as $72$ vulnerabilities
reported. Given such outlying components, the \texttt{Outlier} metric appearing
in the regression specification \eqref{eq: wp ols} is operationalized as a dummy
variable that takes a value one in case a component's reported vulnerabilities
exceed the $0.95$ quantile of vulnerabilities reported; a value zero is used for
all other components. Then: even though the distribution for the counts of
reported and patched vulnerabilities has a similar shape to the counts of all
vulnerabilities reported, many of the components have not patched a single
vulnerability reported. This observation explains the odd-looking shape of the
third histogram in Fig.~\ref{fig: wp vulns}. In other words, about 49\% of the
components have not patched a single vulnerability reported, whereas about 44\%
of the components have patched a single~vulnerability~reported.

The left-hand side plot in Fig.~\ref{fig: wp resup} illustrates the
operationalization of the \texttt{TopReporters} dummy variable; a value one is
recorded for it in case any of the vulnerabilities reported for a component was
made by a productive reporter. Such productive reporters are defined as those
19\% who have reported more than five vulnerabilities. Although the cutoff
point, as visualized by the vertical line in the left-hand side plot, is
arbitrary due to the distribution's shape, \texttt{TopReporters} still serves as
a statistical control analogously to the \texttt{Outlier} dummy
variable. Regarding the right-hand side plot, a little below half of the
components have been updated within a year at the time of the data
collection. However---and due to a long-tail again, the mean is $468$ days and
the maximum as long as $19$ years. The maximum value indicates that a few
outlying components were last updated circa during the early history of
WordPress; the version 1.0 was released in 2004. Even though such outliers may
certainly raise eyebrows among users, including with respect to proper
maintenance and in relation to the \texttt{Removed} metric, it is impossible to
make definite statements because some components may really be finished
products; ``some code really is `done'''~\cite[p.~4]{Cox19}. For instance, a
simple theme released for WordPress 1.0 may work well still today.

Although CVSS base scores are difficult to robustly compare due to changes in
the scoring standard, the leftmost plot in Fig.~\ref{fig: wp cvssstars}
indicates slightly higher severity compared to a replicated
study~\cite{Ruohonen19EASE}. The median scores are also higher than what has
been reported for web applications written in Python~\cite{Ruohonen18IWESEP}. In
particular, there are a few components for which vulnerabilities with the
maximum severity score of 10 have been reported. This observation re-stresses
that vulnerabilities affecting third-party WordPress plugins can be dangerous
too~\cite{Walden10}. A~good example would be the 2025 compromise of the download
site of the Xubuntu Linux distribution that was supposedly done via a vulnerable
third-party plugin~\cite{LWN25}. Regarding the two other plots in Fig.~\ref{fig:
  wp cvssstars}, these serve to add additional evidence about the long-tailed
nature of online star-ratings~\cite{Pajola23, Ramos15}.

\subsubsection{Regression Analysis}\label{subsec: regression analysis 2}

The regression results are shown in Table~\ref{tab: reg 2 qp}. The discussion
can be started by noting a few points from the diagnostics provided. According
to the variance inflation factors (VIFs) reported, multicollinearity is not a
problem in any of the models estimated. Although threshold values for VIFs are
more or less arbitrary~\cite{Obrien07}, the values are clearly below a value of
ten often discussed in the literature~\cite{Prunier15}. The values are also
smaller than a tighter threshold value $2.5$ considered in recent
research~\cite{Cassee25}. However, normality and heteroskedasticity are
problems.

\begin{table*}[th!b]
\centering
\begin{threeparttable}
\caption{Quasi-Poisson Regression Results of the Vulnerability Counts$^1$}
\label{tab: reg 2 qp}
\begin{tabular}{llrcrcr}
\toprule
&\qquad\qquad\qquad& \multicolumn{5}{c}{Vulnerabilities} \\
\cmidrule{3-7}
Metric && All && Patched && Unpatched \\
\hline
\texttt{Installs: [1, 100]} (ref.) \\
\texttt{Installs: [101, 1000]}
&& \e{$0.185$} && \e{$0.216$} && \e{$0.137$} \\
\texttt{Installs: [1001, 10000]}
&& \e{$0.417$} && \e{$0.548$} && \e{$-0.397$} \\
\texttt{Installs: [10001, 100000]}
&& \e{$0.550$} && \e{$0.695$} && \e{$-1.122$} \\
\texttt{Installs: [100001, $\infty$]}
&& \e{$0.754$} && \e{$0.884$} && \e{$-2.029$} \\
\texttt{WordPress: 1-3 series} (ref.) \\
\texttt{WordPress: 4 series}
&& $0.008$ && $0.014$ && $0.056$ \\
\texttt{WordPress: 5 series}
&& $0.032$ && $0.034$ && $-0.070$ \\
\texttt{WordPress: 6-7 series}
&& \e{$0.050$} && $0.043$ && $-0.186$ \\
\texttt{Plugins} (ref.) \\
\texttt{Themes}
&& $-0.093$ && $0.101$ && $0.082$ \\
\texttt{Hosted} (ref.) \\
\texttt{Removed}
&& $0.331$ && $0.375$ && $0.590$ \\
\texttt{OtherReporters} (ref.) \\
\texttt{TopReporters}
&& \e{$-0.102$} && \e{$-0.132$} && \e{$0.506$} \\
\texttt{NonOutlier} (ref.) \\
\texttt{Outlier}
&& \e{$1.711$} && \e{$1.700$} && \e{$2.371$} \\
\hline
\texttt{Reporters}
&& $0.028$ && $0.008$ && \e{$0.267$} \\
\texttt{LastUpdated}
&& \e{$< -0.001$} && \e{$-0.001$} && \e{$< 0.001$} \\
\texttt{Tags}
&& \e{$0.042$} && \e{$0.052$} && $-0.026$ \\
\texttt{CVSS}
&& \e{$0.023$} && \e{$0.028$} && \e{$-0.061$} \\
\texttt{Stars: 4+5}
&& \e{$< 0.001$} && \e{$< 0.001$} && \e{$0.002$} \\
\texttt{Stars: 1+2+3}
&& $< -0.001$ && $< -0.001$ && \e{$-0.002$} \\
\hline
Dispersion parameter
&& $1.139$ && $1.331$ && $0.944$ \\
Maximum VIF value$^2$
&& $1.845$ && $1.721$ && $1.841$ \\
Jarque-Bera \cite{Jarque80} test ($p$-values)$^3$
&& $< 0.001$ && $< 0.001$ && $< 0.001$ \\
Breusch-Pagan \cite{BreuschPagan79} test ($p$-values)$^4$
&& $< 0.001$ && $< 0.001$  && $< 0.001$ \\
Pseudo-R$^3$ of Nagelkerke \cite{Nagelkerke91}$^5$
&& $0.949$ && $0.962$ && $0.392$ \\
Pseudo-R$^4$ of a model without \texttt{Outlier}$^5$
&& $0.703$ && $0.812$ && $0.298$ \\
\bottomrule
\end{tabular}
\begin{tablenotes}
\begin{scriptsize}
\vspace{3pt}
\item{$^1$~The $p$-values are those reported by the\texttt{glm}
  function in R with the \texttt{log} link for the \texttt{quasipoisson} family; values less than $0.05$ are colored.}
\vspace{3pt}
\item{$^2$~Computed with the \texttt{VIF} function from the \texttt{regclass}
  package~\cite{regclass}.}
\vspace{3pt}
\item{$^3$~Computed with the \texttt{jarque.bera.test} function from the \texttt{tseries} package~\cite{tseries}.}
\vspace{3pt}
\item{$^4$~Computed with the \texttt{bptest} function from the \texttt{lmtest} package~\cite{lmtest}.}
\item{$^5$~Computed with the \texttt{NagelkerkeR2} function from the \texttt{fmsb} package~\cite{fmsb}.}
\end{scriptsize}
\end{tablenotes}
\end{threeparttable}
\end{table*}

For each of the six models estimated, the null hypotheses of homoskedasticity
and normality are rejected according to the Breusch-Pagan \cite{BreuschPagan79}
and Jarque-Bera~\cite{Jarque80} tests, respectively. Regarding normality,
particularly the residuals from the models of reported but unpatched
vulnerabilities indicate problems, as could also be expected due to the third
plot in Fig.~\ref{fig: wp vulns}. Otherwise quantile-quantile plots, which are
not explicitly presented for brevity, do not indicate severe problems. When
scatter plots of residuals and estimated values are further examined (but not
explicitly presented), the heteroskedasticity is due to the dummy variables
present in the regression specification \eqref{eq: wp ols}. Therefore, a grain
of salt should be taken when interpreting the results with respect to
statistical~significance.

Turning to the auxiliary hypotheses $\textmd{H}_{21}$, $\textmd{H}_{22}$, and
$\textmd{H}_{23}$; the coefficients for the installation amounts are
statistically significant at the 95\% confidence level and they follow the
expected ordering. The more there are installations online, the more there are
also vulnerabilities reported as well as vulnerabilities reported and
patched. The opposite applies for vulnerabilities reported but unpatched; those
WordPress components having large installation bases tend to have fewer
vulnerabilities unpatched. This observation adds further weight and nuance to
the popularity hypothesis and its relation to software maintenance. In contrast,
the evidence for $\textmd{H}_{24}$ is weak. Although this auxiliary hypothesis
mostly holds on paper, the magnitudes of the $\hat{\beta}_{15}$ and
$\hat{\beta}_{16}$ coefficients are small.

 Finally, and in contrast to the prior precaution raised in
 Subsection~\ref{subsec: wp operationalization}, the overall statistical
 performance is decent. Even without \texttt{Outlier}, the Nagelkerke's
 \cite{Nagelkerke91} pseudo-R$^2$ values are above or equal to zero point seven
 for all vulnerabilities reported and for reported and patched
 vulnerabilities. The much lower pseudo-R$^2$ value for reported but unpatched
 vulnerabilities again exemplifies the problems with this particular vulnerability
 count metric. Nevertheless and all in all, particularly the primary statistical
 effects of interest, including with respect to $\textmd{H}_{22}$~and~
 $\textmd{H}_{23}$, allow to conclude that the popularity hypothesis holds also
 with the second PHP dataset.

\section{Conclusion and Discussion}\label{sec: conclusion}

The paper revisited a popularity hypothesis in empirical software security
research; an assumption that a software's popularity can explain vulnerability
counts reported for the software. According to the results presented, the
hypothesis can be concluded to hold: popular PHP packages and popular WordPress
components have both seen more reported vulnerabilities throughout their release
histories than unpopular packages and components. By implication, the paper
fails to replicate the conclusions of two existing works~\cite{Sakib25,
  Siavvas18} but affirms the conclusion of one existing
work~\cite{Ruohonen19EASE}. The confirmation of the hypothesis has implications
for empirical software security research more broadly. Among these implications
are terminology and construct validity.

It has been argued recently that vulnerability counts should be approached with
care and a terminological adjustment is needed toward \textit{reported}
vulnerabilities~\cite{Meneely25}. The replication results presented support the
argument. In other words, the vulnerabilities reported and observed do not
necessarily tell about software security \textit{per~se} but about something
else, including, but not limited to, incentives and the famous Linus'
law. Whether and how well reported vulnerabilities \textit{may} still proxy
software security remains open to a debate. Given the well-known security risks
involved with public and particularly public full disclosure of software
vulnerabilities~\cite{Ruohonen20CHB}, the question does not necessarily have a
straightforward answer.

There are also implications for empirical research. Recently, it has been argued
that probabilistic random sampling should be preferred in empirical software
engineering to fix a real or perceived generalizability
crisis~\cite{Baltes22}. Analogously to previously solicited expert
opinions~\cite{Marois22}, the results presented cast doubt upon the
argument. The unconditional probability of picking a PHP package without
reported vulnerabilities from the almost entire Packagist population observed is
$0.998$. In other words, it would be very likely that only packages without
reported vulnerabilities would end up in a sample picked randomly. By
implication, it is necessary to again return to the concept of reported
vulnerabilities discussed. Thus, should one consequently believe that the
$381,216$ packages in the sample without reported vulnerabilities are free of
security issues and generally of high quality?  Although no definite answers can
be given, many would likely prefer a negative answer to the question.

The confirmation of the popularity hypothesis has also other implications. As
was noted, GitHub-based popularity metrics have also been used to grant cyber
security funding for open source software projects. As the funding grants have
involved also testing and security audits~\cite{Ruohonen26EnCyCriS26}, it may be
that the evidence for the popularity hypothesis also strengthens in the future
in a sense that even more vulnerabilities are reported for popular open source
software projects. The same point applies with respect to automated tools. For
instance, also large-scale security scanning efforts of open source software
projects have used popularity and related metrics to select preferable samples
for scanning~\cite{AlphaOmega25}. Audits and scanning are important topics
because they are also closer to software security than what is available by
observing reported vulnerabilities.

Finally, both the replicated studies and this replication suffer from a
generalizability problem. That is, it remains unclear whether the hypothesis
would, or would not, hold with a further different dataset. As said, however, it
is also generally unclear how the problem should be addressed---and whether it
is even possible to address it in empirical software security research and
empirical software engineering in general. Regardless, as it stands, there is
now conflicting evidence about the popularity hypothesis. As always with
replication studies \cite[p.~76]{Cockburn20}, it is up to a reader to make the
final verdict about whether popularity affects reported vulnerability counts in
light of the empirical evidence put forward.

\newpage
\balance
\bibliographystyle{acm}

\end{document}